\documentclass[aps,pra,twocolumn,showpacs,superscriptaddress]{revtex4}
\usepackage{graphicx}
\usepackage{amsmath}
\usepackage{amssymb}
\usepackage{lscape}

\input{epsf}

\begin{document}

\title{Few-boson system with a 
single impurity: Universal bound states 
tied to Efimov trimers}

\author{D. Blume}
\affiliation{Homer L. Dodge 
Department of Physics and Astronomy, The University of Oklahoma, 
440 West Brooks Street,
Norman, Oklahoma
73019, USA}

\date{\today}

\begin{abstract}
Small weakly-bound droplets determine a number
of properties of ultracold Bose and Fermi gases.
For example, Efimov trimers near the atom-atom-atom and
atom-dimer thresholds lead to enhanced losses from
bosonic clouds. Generalizations to four- and higher-body
systems have also been considered. Moreover,
Efimov trimers have been predicted to play a role
in the Bose polaron with large boson-impurity scattering length.
Motivated by these considerations,
the present work provides a detailed theoretical analysis of 
weakly-bound $N$-body clusters consisting of $N-1$
identical bosons (denoted by ``B'') of mass $m$ that interact with a single
distinguishable impurity particle (denoted by ``X'') of mass $M$.
The system properties are analyzed as a function of the
mass ratio $\kappa$ (values from $\kappa=1$ to $50$
are considered), where $\kappa$ is equal to $m/M$, and the 
two-body $s$-wave
scattering length 
$a_{\text{BX}}$ between the bosons and the impurity.
To reach the universal Efimov regime in which the size
of the BBX trimer as well as those of larger clusters
is much larger than the
length scales of the underlying interaction
model, three different approaches are 
considered:
resonance states are determined
in the absence of BB and BBX interactions,
bound states are determined in the presence of repulsive
three-body boson-boson-impurity interactions, and
bound states are determined in the presence of
repulsive two-body boson-boson interactions.
The universal regime, in which the details of the underlying interaction model
become irrelevant, 
is identified.
\end{abstract}

\maketitle

\section{Introduction}
\label{sec_introduction}
Ultracold single- and dual-species atomic gases can nowadays 
be prepared and manipulated with exquisite precision.
This has paved the way for the study of various phenomena,
including the Mott-insulator transition~\cite{blochNature}, 
topological defects
such as vortices~\cite{JILA,MIT}, as well as fermionic
and bosonic polarons~\cite{zwierleinFermi,salomonFermi,grimmFermi,JILAbose,AarhusBose}. 
Polarons, which have been studied
extensively in the context of electronic systems, are 
quasi-particles with an effective mass that, typically, differs from
the mass of the underlying 
constituents~\cite{reviewArticles1,reviewArticles2}. 
It has recently 
been proposed that the energy of the ground state Bose polaron at unitarity
is governed by Efimov physics in the low- to medium-density
regime~\cite{PRLMonash,PRXMonash}, i.e., the polaron energy 
is in these regimes predicted to be given by
$- \eta \hbar^2 / (m |a_-|^2)$, where $\eta$ is a dimensionless
universal number and $a_-$ the boson-impurity scattering length at which
the BBX trimer hits the three-atom threshold on the negative
boson-impurity scattering length side.

More specifically, the
equal-mass Bose polaron at unitarity 
was considered within a variational framework~\cite{PRXMonash}.
Treating the Bose polaron using up to two Bogoliubov excitations, it was shown that the 
low-density equation of state is governed by the energy of the BBX Efimov trimer.
Using a more flexible wave function, which allows for up to three Bogoliubov
excitations, the low-density energy is, instead, governed by
the BBBX tetramer that is attached
to the BBX trimer. 
These findings raise two important questions:
Does the inclusion of more Bogoliubov excitations change the equation
of state of the Bose polaron in the low- and medium-density regimes?
Does, and if so how, the picture change if one considers
mass-imbalanced systems?
This paper focuses on the determination of 
weakly-bound few-boson systems with a single impurity.
A good understanding of the hierarchy of few-body states
is a prerequisite for answering the questions raised above.

For single-component
bosons, the properties of the four-body system have been
mapped out in 
detail~\cite{Platter,vonStecher,Deltuva1,Deltuva2}. 
At unitarity, i.e., for an infinitely
large $s$-wave scattering length
(there exists only one scattering length in this case), two four-body states
are universally tied to each Efimov trimer.
In general, the four-body
states are resonance states with finite lifetimes~\cite{Deltuva1,Deltuva2,vonStecher2};
encouragingly,
the lifetimes are sufficiently long for tetramers to be
observed in ultracold gas experiments~\cite{grimm1,grimm2,hulet}.
The properties of these resonance states,
including their convergence to the universal limit,
were studied using a momentum space based formalism~\cite{Deltuva1,Deltuva2}.
The universal limit has 
also been reached---at least in an approximate fashion---by
increasing the size of the lowest Efimov trimer via a repulsive three-body
potential~\cite{vonStecherJPB,kievsky,yan2015}. 
This approach provides approximate values for the
universal energy ratios but not, in general, about 
lifetimes. 

For two-component systems, comparatively little is known about
$N$-body states tied to Efimov 
trimers~\cite{PRLMonash,PRXMonash,Fonseca,BlumeYanPRL,EsryBlakePRL,GiorginiPRA1,wang,hiyama,Naidon,monash1,monash2}. 
Assuming that the impurity and the bosons have the same mass 
but are distinguishable,
the four-body system has been found to display characteristics
that are similar to the single-component case~\cite{PRXMonash}.
Specifically, two tetramer states have been predicted to be tied
to each Efimov trimer on the negative scattering length side. A key difference, though, exists
in the scaling parameter $\lambda^2$, which determines the energy
spacing between consecutive Efimov trimers at unitarity.
This scaling parameter is $\lambda^2=22.7^2$ for the BBB system (identical
particles) and $1986.1^2$ for the BBX system
(assuming equal masses but vanishing BB interactions)~\cite{Efimov1,Efimov2,Efimov3,BraatenReview}.
For $\kappa=8-50$, the energies of the BBBX system,
as well as those of five- and selected six-body systems, 
were determined in Ref.~\cite{BlumeYanPRL}.
The present work extends this earlier Bose-environment-impurity
study in several 
directions:
(i) The mass ratio ``gap'' between 1 and 8 is filled.
(ii) Three different classes of few-body model Hamiltonian are considered and 
their performance with respect to providing 
universal descriptions is compared.
(iii) Selected results for the lifetime of
four-body resonance states
are reported.
(iv) Selected five- and six-particle results are presented.

The remainder of this paper is organized as follows. 
Section~\ref{sec_hamiltonian}
introduces the few-body Hamiltonian models considered while
Secs.~\ref{sec_ecg} and \ref{sec_complexscaling} review the numerical techniques
employed to solve the non-relativistic
time-independent few-particle Schr\"odinger equation.
Results for infinite and negative interspecies
scattering lengths
$a_{\text{BX}}$ are presented in Secs.~\ref{sec_unitarity}
and \ref{sec_negativeas}.
Finally,
summarizing remarks are presented in Sec.~\ref{sec_conclusion}.


\section{System under study and numerical approaches}
\label{sec_theory}


\subsection{System Hamiltonian}
\label{sec_hamiltonian}

We consider $N-1$ identical bosons of mass $m$ 
with position vectors $\vec{r}_j$ ($j=1,\cdots,N-1$) 
interacting with a single
impurity of mass $M$ with position vector $\vec{r}_N$.
Since we consider a single impurity,
its statistics, i.e., whether it is a boson
or fermion, does not play a role.
The mass ratio
$\kappa$,
\begin{eqnarray}
\kappa = \frac{m}{M},
\end{eqnarray}
is varied from 1 to 50. 
The $\kappa \ll 1$ regime was recently investigated
in Refs.~\cite{monash1,monash2}.
Our goal is to describe four- and higher-body states
that are universally linked to BBX Efimov trimers.
This implies that we are considering few-particle Hamiltonian $H$,
for which the magnitude of the $s$-wave scattering length $a_{\text{BX}}$
is large compared to the ranges of the underlying interaction model. 
Moreover, the
size of the Efimov trimer should be much larger
than the ranges of the underlying interactions.

The few-particle Hamiltonian $H$
accounts for the kinetic energy of each of the particles,
a two-body interaction potential $V_{\text{BX}}(r_{jN})$
for the BX pairs, 
a two-body interaction potential $V_{\text{BB}}(r_{jk})$
for the BB pairs, 
and a three-body potential $V_{\text{BBX}}(r_{jk},r_{jN},r_{kN})$
for the BBX triples, 
\begin{eqnarray}
\label{eq_ham}
H = 
-\frac{\hbar^2}{2m} \sum_{j=1}^{N-1} 
\nabla^2_{\vec{r}_j}
-
\frac{\hbar^2}{2M} \nabla^2_{\vec{r}_N}
+
\sum_{j=1}^{N-1} V_{\text{BX}}(r_{jN})
+ \nonumber \\
\sum_{j=1}^{N-2} \sum_{k>j}^{N-1} V_{\text{BB}}(r_{jk})
+
\sum_{j=1}^{N-2} \sum_{k>j}^{N-1} V_{\text{BBX}}(r_{jk},r_{jN},r_{kN}).
\end{eqnarray}
The distances $r_{jk}$ are defined through $r_{jk}=|\vec{r}_j-\vec{r}_k|$.
Throughout we treat the two-body
$s$-wave scattering length $a_{\text{BX}}$
of the BX pairs as a tunable parameter. This is accomplished
by changing the depth $d_{\text{BX}}$ of a purely attractive two-body Gaussian
potential while keeping the range $r_{\text{BX}}$
constant,
\begin{eqnarray}
V_{\text{BX}}(r_{jN})= d_{\text{BX}} \exp \left[-
\frac{(r_{jN})^2}{2 (r_{\text{BX}})^2}
\right].
\end{eqnarray}
The depth $d_{\text{BX}}$ ($d_{\text{BX}}<0$)
is restricted to values for which $V_{\text{BX}}(r_{jN})$
supports at most a single two-body $s$-wave bound state
in free space. 
This implies that we eliminate a large set of ``high-energy channels''
from the outset.
As will become clear below, our model Hamiltonian
also excludes weakly- and deeply-bound
BB molecules.
The unitary point, where $a_{\text{BX}}$ diverges (i.e.,
where $a_{\text{BX}}$ is infinitely large),
is of particular interest in this work. At unitarity, the two-body binding
energy vanishes and the two-body interaction is,
in the $r_{\text{BX}} \rightarrow 0$ limit, not characterized by a length
scale. Throughout, we consider finite two-body ranges $r_{\text{BX}}$.
For our results to be universal, it is necessary to work in 
the parameter regime where the
sizes of the dimers, trimers, and larger clusters are much larger than
the range $r_{\text{BX}}$.
Note that our interaction is single-channel in nature
and that universality refers to zero-range
universality and not van der Waals universality~\cite{grimm,greene,naidon1,naidon2}.

The BB interaction potential $V_{\text{BB}}(r_{jk})$ is 
also modeled by a Gaussian potential,
\begin{eqnarray}
V_{\text{BB}}(r_{jk})= d_{\text{BB}} \exp 
\left[ -
\frac{(r_{jk})^2}{2 (r_{\text{BB}})^2}
\right].
\end{eqnarray}
In contrast to the BX potential, which is purely attractive, the
BB potential is chosen to vanish or to be purely repulsive
with a positive BB $s$-wave scattering
length $a_{\text{BB}}$.
Even though the use of a purely repulsive interaction potential
is unphysical
(typical van der Waals potentials
relevant to cold alkali gases have,
regardless of the sign of the $s$-wave scattering length,
an attractive pocket),
the model should yield reasonable results
provided the  
BB scattering length is much smaller than
the magnitude of the BX scattering length, i.e., for
$a_{\text{BB}} \ll |a_{\text{BX}}|$.

Lastly, the three-body interaction $V_{\text{BBX}}$ is parametrized
via a purely repulsive Gaussian potential with barrier $d_{\text{BBX}}$
($d_{\text{BBX}} \ge 0$)
and range $r_{\text{BBX}}$,
\begin{eqnarray}
V_{\text{BBX}}(r_{jk},r_{jN},r_{kN})= \nonumber \\
d_{\text{BBX}} 
\exp
\left[-
\frac{(r_{jk})^2 + (r_{jN})^2 + (r_{kN})^2}{
2 (r_{\text{BBX}})^2}
\right].
\end{eqnarray}
The use of a repulsive three-body potential facilitates reaching the regime
where the bound states of 
the three- and higher-body clusters are large compared
to the length scales of the underlying interaction 
potentials~\cite{vonStecherJPB,kievsky,yan2015,BlumeYanPRL}.
Specifically, a non-zero $d_{\text{BBX}}$ 
can lead to a large BBX ground state
trimer, which mimicks the 
behavior of large, universal excited Efimov trimer states.
If we consider the case where 
$d_{\text{BX}}<0$
and $d_{\text{BB}}=0$, then
the three-body potential $V_{\text{BBX}}$
can be interpreted as setting the value of the three-body
parameter. It was shown in Ref.~\cite{BlumeYanPRL} 
for $\kappa=8-50$ that the BBBX
ground state energies are, if expressed in units of the BBX 
ground state energies,
to a good 
approximation independent of the value of $d_{\text{BBX}}$ provided
$d_{\text{BBX}}$ is,
for constant $r_{\text{BBX}}$, sufficiently large.
For small $d_{\text{BBX}}$,
in contrast,
the three-body potential serves as a perturbation that modifies 
the, in general, non-universal ground states of the Hamiltonian with 
$d_{\text{BX}}<0$
and $d_{\text{BB}}=0$.

The model Hamiltonian $H$, Eq.~(\ref{eq_ham}), has a large number of
parameters: the mass ratio $\kappa$;
the ranges $r_{\text{BX}}$, $r_{\text{BB}}$, and $r_{\text{BBX}}$;
the BX and BB scattering lengths $a_{\text{BX}}$ and
$a_{\text{BB}}$
(or, alternatively, the parameters $d_{\text{BX}}$
and $d_{\text{BB}}$); 
and the strength $d_{\text{BBX}}$ of the three-body potential.
Given the large number of parameters, we cannot exhaustively
explore the complete parameter space.
Our non-exhaustive study
considers three different sub-classes of the Hamiltonian $H$,
referred to as Model~I--Model~III:
\begin{itemize}
\item Model~I: $d_{\text{BX}}<0$, $d_{\text{BB}}=0$, and
  $d_{\text{BBX}}=0$.
\item Model~II: $d_{\text{BX}}<0$, $d_{\text{BB}}=0$, and
  $d_{\text{BBX}}>0$.
\item Model~III: $d_{\text{BX}}<0$, $d_{\text{BB}}>0$, and
  $d_{\text{BBX}}=0$.
\end{itemize}
The ground states and likely also a subset of the 
excited eigen states supported by Model~I are
expected to be ``contaminated'' by, possibly significant,
finite-range or non-universal corrections. 
Sufficiently high in the energy spectrum, however, the 
three-body bound states supported by Model~I
exhibit Efimov characteristics and the associated
four-body resonance states should exhibit model-independent 
properties.
For Models~II--III, we calculate bound states but not resonance states.
The premise is that the repulsive BBX and BB potentials serve to 
push the particles out, leading---for certain parameter 
combinations---to ground states that are large
compared to the length scales of the underlying interaction potentials.
The BBX energies 
depend on the parameters of the
Hamiltonian model.
However, universality implies that the B$_{N-1}$X energies
for $N \ge 3$, if measured in units of one of the BBX Efimov trimer
energies, are 
independent of the details of the underlying model Hamiltonian.

A key goal of this work is to determine universal energy ratios 
for few-body systems as a function of the mass ratio and
to illustrate convergence toward these universal energy ratios for 
the different models.
Model~I was employed in Ref.~\cite{EsryBlakePRL} for large mass ratios,
Model~II in Ref.~\cite{BlumeYanPRL} for $\kappa=8-50$, and
a model similar to Model~III
in Ref.~\cite{GiorginiPRA1,GiorginiPRA2} for systems with equal masses and 
relatively small mass
imbalance,
with the BX and BB
Gaussian potentials replaced by square well potentials.

Throughout, we set $r_{\text{BX}}=r_{\text{BB}}=r_{\text{BBX}}/\sqrt{8}$
and vary $d_{\text{BX}}$, $d_{\text{BB}}$, and $d_{\text{BBX}}$.
We use $r_{\text{BX}}$ to define the short-range energy scale
$E_{\text{sr}}$,
\begin{eqnarray}
  E_{\text{sr}} = \frac{\hbar^2}{2
    \mu r_{\text{BX}}^2},
\end{eqnarray}
where the two-body reduced mass $\mu$ is defined as
$\mu=mM/(m+M)$.


\subsection{Determination of bound states}
\label{sec_ecg}
The few-body bound states considered in this work 
have vanishing total relative orbital angular momentum $L$
and positive
relative parity $\Pi$. To determine the $L^{\Pi}=0^+$
bound state energies, we 
separate off the three 
center of mass degrees of freedom 
(the relative Hamiltonian is denoted by $H_{\text{rel}}$) and
solve the relative Schr\"odinger equation
\begin{eqnarray}
H_{\text{rel}} \psi=E \psi
\end{eqnarray}
by expanding the eigen states $\psi$ in
terms of explicitly correlated Gaussian basis functions 
$\phi_l$~\cite{CGbook,VargaRMP},
\begin{eqnarray}
\label{eq_wavefunction}
\psi  = \sum_{l=1}^{N_b} c_l 
{\cal{S}}(\phi_l(\vec{X})),
\end{eqnarray}
where
\begin{eqnarray}
\label{eq_basisfunction}
\phi_l(\vec{X})=
\exp \left( -\frac{1}{2} \vec{X}^T \underline{A}_l \vec{X} \right).
\end{eqnarray}
Here, 
$\vec{X}$ collectively denotes 
a set of $N-1$ relative Jacobi vectors,
$N_b$ the number of unsymmetrized
basis functions, and $\underline{A}_l$ a 
$(N-1) \times (N-1)$ parameter matrix.
The linear parameters $c_l$ are obtained by diagonalizing the
generalized eigen value problem spanned by the
relative Hamiltonian matrix $\underline{H}_{\text{rel}}$
and
the overlap matrix $\underline{O}$, 
whose $ll'$ element is given by 
$\langle \phi_l|\phi_{l'} \rangle$. 
The overlap matrix enters since the basis functions
are not
orthogonal to each other. 
Importantly, all matrix elements have compact
analytical expressions. The $N(N-1)/2$ non-linear variational
parameters contained in the symmetric $\underline{A}_l$ matrices
are determined through a semi-stochastic optimization
procedure~\cite{kukulin}. 
In Eq.~(\ref{eq_wavefunction}),
${\cal{S}}$ denotes a symmetrizer, which ensures
that the basis functions are symmetric under the exchange of the
position vectors of any two 
identical bosons. 

The explicitly correlated Gaussian basis set expansion approach 
has several characteristics that make their use advantageous in
the context of Efimov studies. The non-linear variational
parameters can be chosen to describe different ``geometries''
such as a ``3+1 configuration'', where one atom is very loosely bound
to a more tightly bound trimer~\cite{BlumeYanPRL}. 
Moreover, since the basis
set is constructed using non-orthogonal
basis functions that cover vastly different
length scales, bound states whose
sizes range from the two-body ranges $r_{\text{BX}}$
and $r_{\text{BB}}$ to several 10 or 100 times $r_{\text{BX}}$
and $r_{\text{BB}}$ can be generated~\cite{BlumeYanPRL}.
Another useful feature is that one can construct separate
basis sets for each of the eigen states. This has the benefit
that the basis set can be targeted toward a specific state
and that a comparatively small basis set may provide
an excellent description of a given eigen state~\cite{RakshitPRA}.


\subsection{Determination of resonance states}
\label{sec_complexscaling}

The states supported by 
$H_{\text{rel}}$ can be grouped into three classes:
(i) bound states, which are characterized by an
exponentially decaying tail at large distance scales;
(ii) scattering states, which display oscillatory behavior 
in one or more distance coordinates; and
(iii) resonance states, which are characterized
by exponential
growth in at least one of the distance coordinates.
As discussed in what follows,
the explicitly correlated Gaussian approach
can be generalized to treat resonance states via the 
complex scaling approach~\cite{GeneralRef1,GeneralRef2}.
The $l$-th basis function given in Eq.~(\ref{eq_basisfunction})
can be rewritten as~\cite{CGbook}
\begin{eqnarray}
\label{eq_basisfunction2}
\phi_l(\vec{X})=
\exp \left[ -\sum_{j<k}^N \frac{r_{jk}^2}{2 (\alpha_{l,jk})^2} \right],
\end{eqnarray}
where the non-linear width parameters $\alpha_{l,jk}$ are determined
by the elements $a_{l,jk}$ of the matrix
$\underline{A}_l$.
Equation~(\ref{eq_basisfunction2}) shows  that
the basis functions fall off exponentially
as one or more of the interparticle distances become large.
This illustrates that the basis functions cannot be used
(at least not directly) to expand resonance states, which
contain an exponentially growing piece.
In general, this holds true for nearly all basis functions
that are designed 
to describe bound states of hermitian Hamiltonian~\cite{VargaRMP}.

The complex scaling approach provides a means to use
basis functions such as those given in Eq.~(\ref{eq_basisfunction})
to describe resonance states~\cite{VargaRMP,GeneralRef1,GeneralRef2,bromley,jonsell}.
To this end, the vector
$\vec{X}$ is rotated into the complex plane~\cite{GeneralRef1,GeneralRef2},
\begin{eqnarray}
\vec{X} ' = 
U \vec{X},
\end{eqnarray}
where 
$U$ is equal to
$\exp( \imath \theta)$ and
 $\theta$ is an appropriately chosen rotation angle.
The transformed Schr\"odinger equation reads
\begin{eqnarray}
\label{eq_setilde}
\tilde{H}_{\text{rel}} \tilde{\psi} = \tilde{E} \tilde{\psi},
\end{eqnarray}
where $\tilde{H}_{\text{rel}}= U^{\dagger} H_{\text{rel}} U$ 
and $\tilde{\psi}= U^{\dagger} \psi$.
To find the eigen energies $\tilde{E}$, we 
expand
\begin{eqnarray}
\tilde{\psi}  = \sum_{l=1}^{N_b} d_l {\cal{S}} (\phi_l(\vec{X})),
\end{eqnarray}
where the $\phi_l(\vec{X})$ are defined in Eq.~(\ref{eq_basisfunction})
and where the $d_l$ are complex (linear)
expansion coefficients. 
Since the matrix elements $(\tilde{H}_{\text{rel}})_{ll'}$ are complex, the generalized
eigen value problem is spanned by the complex  Hamiltonian
matrix $\tilde{\underline{H}}_{\text{rel}}$ 
and the real overlap matrix
$\tilde{\underline{O}}$.
The Hamiltonian matrix $\tilde{\underline{H}}_{\text{rel}}$ depends
on
the rotation angle 
$\theta$ but $\tilde{\underline{O}}$ does not
(in fact, we have $\tilde{\underline{O}}=\underline{O}$).
The kinetic energy
contribution to the matrix elements contains an overall factor
of $\exp(2 \imath \theta)$, which can be calculated upfront
for each
$\theta$ considered~\cite{GeneralRef1,GeneralRef2}.
The calculation of the potential energy contribution, in contrast, is more 
involved~\cite{VargaRMP}. 
Since the rotation introduces a $\theta$ dependence in the exponent
of the Gaussian interaction potentials, the 
potential energy contribution to the 
Hamiltonian matrix element has to be calculated separately
for each rotation angle and matrix element. While this is technically
straightforward, it does increase the computational effort compared to the
bound state calculations, especially if a fine resolution in the rotation
angle is desired.

For the basis functions considered here (and more generally,
for all square integrable basis functions), it can be shown,
assuming one has a complete basis set, that
(i) the energies of bound states are independent
of the rotation angle, i.e., $\tilde{E}=E$ for true
bound states;
(ii) the energies of scattering states rotate with the
rotation angle, i.e., $\tilde{E}=\exp(i \theta) E$ for
scattering states; and
(iii) the energies of resonance states live in the
complex plane and are independent of the
rotation angle~\cite{GeneralRef1,GeneralRef2}.
In practice, there tends to exist a limited range
of angles for which the energy $\tilde{E}$ 
does not move in the complex energy plane (is stationary).
The challenge is thus to 
generate a basis set 
for which the energy $\tilde{E}$ is, for a range of rotation angles,
stationary (or stationary within some tolerance). To the best of our knowledge,
a unique approach that accomplishes this does
not exist. The reason is that the variational principle, 
which provides the backbone for most basis set construction
schemes that are aimed at describing bound states,
does not apply to resonance states.

Following the strategy that has been used 
to describe three-particle systems~\cite{bromley,jonsell},
our calculations consist of two steps.
First, we generate a basis set by 
minimizing the energy of a ``target state'' by diagonalizing the
generalized eigen value problem spanned by $\underline{H}_{\text{rel}}$ and 
$\underline{O}$. Specifically, the basis set is increased one basis 
function at a time, with the newly added basis function 
chosen such that the energy of the state whose energy is higher than
but closest to a preset ``target energy'' $E_{\text{target}}$ is minimized.
The target energy is chosen based on the real part of the energy
of the resonance state. If the real part is expected to be
$E_{r}$
(this expectation may derive from previous 
calculations or physics arguments), 
we choose $E_{\text{target}}$ to be comparable
to but above
$E_{r}$.
The calculations are repeated
for different $E_{\text{target}}$ to eliminate 
a possible bias due to the choice of
the actual value of the target energy.
Second, we rotate the basis functions of the basis set
constructed in the first step
and solve the generalized eigen value problem spanned
by $\tilde{\underline{H}}_{\text{rel}}$ and $\tilde{\underline{O}}$
for various angles $\theta$ (typically of order
50-75), where $\theta$ ranges
from $0$ to $0.48$radians ($\theta$
has to be smaller than $\pi/2$).
Importantly, the rotation approach results in the
energy $E_r$ of the resonance state as well as its lifetime
$\tau$,
\begin{eqnarray}
\tau = \frac{\hbar}{2|E_i|};
\end{eqnarray}
throughout, we write the resonance energy
as $E=E_r + \imath E_i$,
where $E_i$ is negative.
The complex scaling approach 
is illustrated in
Appendix~\ref{appendix_numerics}.


\section{Results: Unitarity}
\label{sec_unitarity}
This section presents few-body energies 
for Models~I-III with infinitely large
$s$-wave scattering length $a_{\text{BX}}$.

\subsection{Model~I}

Table~\ref{table_modelI} reports selected bound state energies for
Model~I, which is characterized by a vanishing BB
interaction potential, for $N=3-6$.
In the limit that the trimer size is much larger than the (effective)
range of the BX interaction potential, the $N=3$ energies for Model~I should approach
Efimov's zero-range results.
The second
column of Table~\ref{table_modelI} 
shows that the energy ratio $|E_3^{\text{gr}}|/E_{\text{sr}}$ increases with increasing
mass ratio $\kappa$.
This suggests that the
three-body ground state energies of Model~I are contaminated the most by non-universal
corrections for large mass ratios $\kappa$.
Consistent with
the literature, the energy ratio between
two consecutive three-body energies approaches
the universal zero-range value $\lambda^2$ (see Table~\ref{table_zr})
for sufficiently high excitations. For $\kappa=50$, e.g., the energy ratio
$E_3^{\text{exc},1}/E_3^{\text{exc},2}$ deviates by about $7.6$~\%
from the universal value while the energy ratio
$E_3^{\text{exc},2}/E_3^{\text{exc},3}$
deviates by only about $0.4$~\% 
from the universal value.

\begin{figure}
\vspace*{+.9cm}
\includegraphics[angle=0,width=70mm]{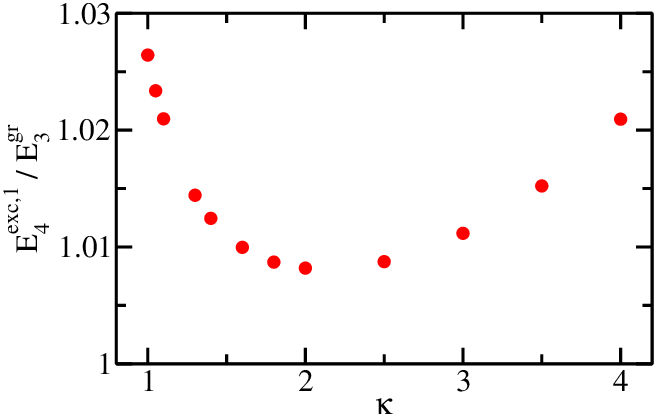}
\vspace*{0.2cm}
\caption{(color online)
Energy of the four-body excited state supported by
Model~I at unitarity.
Circles show the energy
ratio $E_4^{\text{exc},1}/E_3^{\text{gr}}$ 
as a function of the mass ratio $\kappa$.
}
\label{fig_gauss4exc}
\end{figure}

Table~\ref{table_modelI} reveals three trends for $N\ge 4$:
(i) The energy ratios $E_{N}^{\text{gr}}/E_3^{\text{gr}}$ for $N=4-6$
decrease monotonically with increasing $\kappa$.
(ii) The number of four-body bound states increases with increasing $\kappa$.
While we cannot rule out the existence of extremely weakly-bound
four-body
states beyond those reported in Table~\ref{table_modelI}
(our approach yields variational upper
bounds and it is possible that weakly-bound states are not captured by the
basis sets considered), the trend that $N \ge 4$ systems with larger $\kappa$,
described by Model~I,
support more bound states than systems with smaller $\kappa$ is evident.
(iii) The ratio
$E_4^{\text{exc},1}/E_3^{\text{gr}}$ changes,
as also illustrated in Fig.~\ref{fig_gauss4exc}, non-monotonically
with increasing $\kappa$.
The energy ratio takes a minimum at $\kappa \approx 2$ and increases
for both smaller and larger mass ratios
(we explored the regime $1 \le \kappa \le 50$).
We note that the non-monotonic change of the energy ratio
$E_4^{\text{exc},1}/E_3^{\text{gr}}$
with $\kappa$ may be sensitive
to the specifics of the two-body interaction considered. 

\begin{widetext}

  \begin{table}
\caption{
Bound state energies for the B$_{N-1}$X system, $N=3-6$, 
with infinitely large
$s$-wave scattering length $a_{\text{BX}}$ 
and $d_{\text{BB}}=d_{\text{BBX}}=0$
(Model~I) for various mass ratios $\kappa$.
Since $|E_3^{\text{gr}}|/E_{\text{sr}}$ (see column 2)
increases with increasing $\kappa$, the results are expected
to be less universal for larger $\kappa$ than for smaller $\kappa$
(see text for details).
The ``missing entries'' correspond to parameter combinations
where either the calculation
was not attempted or no bound state was found.
The energy ratios in columns 3-11
have uncertainties in the last digit reported.
}
\begin{ruledtabular}
\begin{tabular}{ll|lll|llll|ll}
$\kappa$ & $|E_3^{\text{gr}}|/E_{\text{sr}}$ & $E_3^{\text{gr}}/E_3^{\text{exc,1}}$ & $E_3^{\text{exc,1}}/E_3^{\text{exc,2}}$ &
$E_3^{\text{exc,2}}/E_3^{\text{exc,3}}$ &
$E_4^{\text{gr}}/E_3^{\text{gr}}$ &  $E_4^{\text{exc,1}}/E_3^{\text{gr}}$ &  
$E_4^{\text{exc,2}}/E_3^{\text{gr}}$ &  $E_4^{\text{exc,3}}/E_3^{\text{gr}}$ &  
$E_5^{\text{gr}}/E_3^{\text{gr}}$ &  $E_6^{\text{gr}}/E_3^{\text{gr}}$ \\
\hline
$1$ & $2.354 \times 10^{-4}$ & & & & $12.21$ & $1.026$ & & & $43.60$ & $97.92$ \\
$2$ & $3.470 \times 10^{-3}$ & $23860.$ && & $6.956$  & $1.008$ & & & $18.64$ & $35.20$ \\
$3$ & $9.660 \times 10^{-3}$ & $3398.$ &  && $5.646$ & $1.011$ &&& $13.69$ & $24.35$ \\
$4$ & $1.694 \times 10^{-2}$ & $1140.$ && & $5.046$  & $1.021$ & & & $11.61$ & $20.03$ \\
$8$ & $4.437 \times 10^{-2}$ & $151.8$ & && $4.184$  & $1.130$ & & & $8.840$ & $14.51$ \\
$12$& $6.512 \times 10^{-2}$ & $58.52$ & $66.09$ && $3.892$  & $1.310$ & & & $7.964$ & $12.83$ \\
$16$ & $8.082 \times 10^{-2}$ & $30.98$ & $39.35$ && $3.740$  & $1.489$ & & & $7.519$ & $11.99$ \\
$133/6$ & $9.881 \times 10^{-2}$ & $15.47$ & $23.49$ & $23.67$ & $3.606$  & $1.711$ & & & $7.134$ & $11.27$ \\
$26$ & $1.075 \times 10^{-1}$ & $11.22$ & $18.70$ & $18.90$ & $3.552$ & $1.819$ & $1.013$ && $6.981$ & $10.98$ \\
$30$ & $1.152 \times 10^{-1}$ & $8.575$ & $15.41$ & $15.64$ & $3.508$  & $1.911$ & $1.059$ & & $6.859$ & $10.76$ \\
$35$ & $1.232 \times 10^{-1}$ & $6.587$ & $12.62$ & $12.91$ & $3.467$ & $2.005$ & $1.134$ && $6.742$ & $10.54$ \\
$40$ & $1.300 \times 10^{-1}$ & $5.371$ & $10.66$ & $11.04$ & $3.434$  & $2.082$ & $1.214$ & & $6.651$ & $10.37$ \\
$45$ & $1.358 \times 10^{-1}$ & $4.572$ & $9.182$ & $9.680$ & $3.408$ & $2.145$ & $1.290$ && $6.578$ & $10.24$ \\
$50$ & $1.409 \times 10^{-1}$ & $4.016$ & $8.021$ & $8.648$ & $3.386$  & $2.199$ & $1.362$ & $1.003$ & $6.518$ & $10.13$ \\
\end{tabular}
\end{ruledtabular}
\label{table_modelI}
\end{table}

\end{widetext}

\begin{table}
\caption{
Column~2 shows the scaling parameter $\lambda^2$
predicted by the zero-range theory
for the three-body system for various $\kappa$.
}
\begin{ruledtabular}
\begin{tabular}{ll}
$\kappa$ 
& $\lambda^2$  \\
\hline
$1$ & $(1986.1)^2=3.9447 \times 10^6$ \\
$2$ & $(153.84)^2=23666.$ \\
$3$ & $(57.876)^2=3349.6$ \\
$4$ & $(33.491)^2=1121.6$ \\
$8$ & $(12.488)^2=155.94$ \\
$12$ & $(8.1305)^2=66.105$ \\
$16$ & $(6.2804)^2=39.443$ \\
$133/6$ & $(4.8651)^2=23.670$ \\
$26$ & $(4.3477)^2=18.902$ \\
$30$ & $(3.9553)^2=15.644$ \\
$35$ & $(3.5944)^2=12.920$ \\
$40$ & $(3.3249)^2=11.055$ \\
$45$ & $(3.1152)^2=9.7047$  \\
$50$ & $(2.9470)^2=8.6847$  
\end{tabular}
\end{ruledtabular}
\label{table_zr}
\end{table}

\begin{widetext}

\begin{table}
\caption{
Resonance energies for the B$_{N-1}$X system, $N=4$, interacting through
a BX Gaussian potential with infinitely large
$s$-wave scattering length 
$a_{\text{BX}}$ and $d_{\text{BB}}=d_{\text{BBX}}=0$
(Model~I) for various mass ratios $\kappa$.
Columns~2-5 report four-body resonances tied to the first excited three-body state
with energy $E_3^{\text{exc},1}$,
columns~6-7 report four-body resonances tied to the second excited three-body state
with energy $E_3^{\text{exc},2}$,
and
column~8 
reports four-body resonances tied to the third excited three-body state
with energy $E_3^{\text{exc},3}$.
The missing entries indicate that a calculation was either not attempted or
did not yield reliable results.
In some cases, we were able to determine $E_r$ approximately but not
$E_i$; in these cases, the $E_i$ entry is marked by ``?''.
}
\begin{ruledtabular}
\begin{tabular}{l|llll|ll|l}
$\kappa$ & 
$\left(\frac{E_r}{E_3^{\text{exc,1}}},\frac{E_i}{E_3^{\text{exc,1}}}\right)$ & 
$\left(\frac{E_r}{E_3^{\text{exc,1}}},\frac{E_i}{E_3^{\text{exc,1}}}\right)$ & 
  $\left(\frac{E_r}{E_3^{\text{exc,1}}},\frac{E_i}{E_3^{\text{exc,1}}}\right)$ &
$\left(\frac{E_r}{E_3^{\text{exc,1}}},\frac{E_i}{E_3^{\text{exc,1}}}\right)$ &
$\left(\frac{E_r}{E_3^{\text{exc,2}}},\frac{E_i}{E_3^{\text{exc,2}}}\right)$ & 
$\left(\frac{E_r}{E_3^{\text{exc,2}}},\frac{E_i}{E_3^{\text{exc,2}}}\right)$ & 
  $\left(\frac{E_r}{E_3^{\text{exc,3}}},\frac{E_i}{E_3^{\text{exc,3}}}\right)$
  
\\
\hline
$4$ & $(3.28,\approx 0.03)$ & & & &&& \\
$8$ & $(2.63,0.018)$ & &&&&& \\
$12$ & $(2.35,0.019)$ & & &  &&& \\
$16$ & $(2.19,0.011)$ & && & $(2.22,\approx 0.01)$ && \\
$133/6$ & $(2.19,0.009)$ & $(1.03,0.004)$ & && $(2.08,0.020)$ && $(1.96,?)$ \\
$26$ & $(2.45,0.023)$ & $(1.19,0.028)$ &&& $(2.05,0.016)$ &&   \\
$30$ & $(2.95,0.047)$ & $(1.44,0.047)$ & && $(2.06,0.005)$ & $(1.04,0.003)$ & $(2.00,?)$  \\
$35$ & $(3.46,0.048)$ & $(1.61,0.055)$ & && $(1.93,0.003)$ & $(1.03,0.0006)$ &  \\
$40$ & $(3.73,0.039)$ & $(1.73,0.056)$ & $(1.02,0.010)$ && $(\approx 1.94,?)$ & $(\approx 1.03,?)$ & $(\approx 2.01,?)$  \\
$45$ & $(3.85,0.030)$ & $(1.83,0.051)$ & $(1.20,0.0009)$ & $(1.04,0.015)$ & $(1.86,0.078)$ & $(1.04,?)$ & $(2.07,?)$ \\
$50$ & $(3.90,0.026)$ & $(1.94,0.049)$ & $(1.46,?)$ & $(1.10,0.021)$ & $(1.85,?)$ & $(1.03,?)$ & $(2.40,\approx 0.036)$ \\
\end{tabular}
\end{ruledtabular}
\label{table_modelIres}
\end{table}

\end{widetext}

Analysis of the four-body resonance states that are
tied to $E_3^{\text{exc},j}$ shows, as we will discuss now, that the 
four-body spectra reported in Table~\ref{table_modelI},
especially for large $\kappa$,
are not universal; this is, of course, not
surprising given the discussion presented in Sec.~\ref{sec_theory}.
Table~\ref{table_modelIres} summarizes
the real and imaginary parts $E_r$ and $E_i$
of the four-body resonance energies
for $\kappa = 4-50$.
In Table~\ref{table_modelIres}, $E_r$ and $E_i$ are
reported in terms of the excited three-body
bound state energies $E_3^{\text{exc},j}$.
As mentioned in Appendix~\ref{appendix_numerics},
a precise and unambiguous identification of resonance
states becomes numerically more
challenging as $|E_r|/E_{\text{sr}}$ and/or
$|E_i|/E_{\text{sr}}$
decrease.
Consequently, Table~\ref{table_modelIres} reports results
for resonances tied to three different three-body states for $\kappa=50$ but
only one three-body state for $\kappa=4$.
For $\kappa \le 4$, the complex scaling approach, as implemented
by us,
did not yield reliable four-body results.

We first discuss our results for $\kappa=4-16$.
For these $\kappa$, the ratio $E_r/E_3^{\text{exc},1}$
deviates notably from both the energy ratios
$E_4^{\text{gr}}/E_3^{\text{gr}}$ and
$E_4^{\text{exc,1}}/E_3^{\text{gr}}$, indicating that the
four-body results reported in Table~\ref{table_modelI}
are not universal.
For $\kappa=16$, we were able to reliably determine
$E_r$ for a four-body resonance tied to the second excited
trimer state,
yielding $E_r/E_3^{\text{exc,2}}=2.22$.
Since this value is close to the ratio of $2.19$
obtained for the resonance attached to the first excited trimer,
we conclude that the
energy ratios for the four-body resonances for $\kappa=4-16$, 
tied to the first excited three-body state, are close
to universal.

For larger mass ratios, the four-body resonances tied to the first excited trimer 
are not universal. However, closer to universal 
results 
are obtained  for the resonances
that are tied to the second or third excited trimers.
For $\kappa=30-50$, our complex scaling results suggest that there are two
four-body states tied to each Efimov trimer, with the second state
having a 
resonance position that is just a bit below the corresponding trimer energy.
For smaller $\kappa$, ``excited'' four-body resonance states
with real parts $E_r$
very close to the Efimov trimer energy may also exist.
However, we were not able to describe such resonance states by our approach.
We  note that the identification of the four-body resonances
for $\kappa \gtrsim 35$ is
challenging due to the existence
of multiple four-body resonances.
For $\kappa=35$, e.g., we find a resonance at $E_r \approx 4.9 E_3^{\text{exc},2}$
that is not reported in Table~\ref{table_modelIres}
since we believe that this resonance would not ``survive''
if we went to resonances that are
attached to more highly-excited three-body
states.

Importantly,
the complex scaling calculations also
provide estimates of the lifetimes $\tau$.
If expressed, as in Table~\ref{table_modelIres}, in terms of the
corresponding trimer energies, the imaginary parts $E_i$ 
of the resonance energies are comparable,
in terms of the order of magnitude, to those found for the equal-mass four-boson
system. 
For example, Ref.~\cite{Deltuva2} found 
$E_r = 4.6108 E_{3}^{\text{exc},j}$ 
and $E_i = 0.01484 E_{3}^{\text{exc},j}$ 
for the energetically lower-lying BBBB state
and
$E_r = 1.00228 E_{3}^{\text{exc},j}$ 
and $E_i = 2.38 \times 10^{-4} E_{3}^{\text{exc},j}$ 
for the energetically higher-lying BBBB state
in the large $j$ limit. 
This suggests that signatures 
of the four-body resonance states
of unequal-mass systems should be observable experimentally.

\subsection{Model~II}

Since resonance states are, in general, more challenging to
determine than bound states, it is desirable to 
employ an
interaction model for which the ground state of the trimer behaves
close to universal.
This section summarizes our energies at unitarity for Model~II,
for which the 
repulsive BBX potential leads to a significant
reduction of the binding energy of the 
ground state trimer.
Table~\ref{table_modelII} summarizes three-, four-, and five-body
energies, which are 
obtained for such a large $d_{\text{BBX}}$
that the difference to the infinity limit is rather small
(see also Ref.~\cite{BlumeYanPRL}).
In general, the resulting energy ratios could
depend on the details of the underlying potential model.
For the three-body sector,
we believe that the results reported in Table~\ref{table_modelII} are,
to a very good approximation, universal since
$E_3^{\text{gr}}/E_3^{\text{exc,1}}$
is close to the zero-range prediction for $\lambda^2$.

As discussed in Ref.~\cite{BlumeYanPRL},
the four-body systems 
with $\kappa \ge 16$ support two
four-body states.
One four-body state is roughly twice as strongly
bound as the trimer while the other is extremely weakly bound.
As the mass ratio decreases, the weakly-bound state disappears
(or at least our calculations were not able to describe
it) while the deeper-lying four-body state
becomes more strongly bound.
For $\kappa=1$, e.g., the binding energy of the 
ground state tetramer is roughly 10 times larger
than that of the ground state trimer. In terms of size,
this suggests that the ground state tetramer is 
smaller by about a factor of $\sqrt{10}$
than the ground state trimer. Since
$|E_4^{\text{gr}}|/E_{\text{sr}}$ is still much smaller than 1, we believe that
the tetramer energy is close to universal.
This is confirmed by the fact that Refs.~\cite{PRXMonash,GiorginiPRA1}
found similar ratios for $E_4^{\text{gr}}/E_3^{\text{gr}}$,
namely
$9.35-9.7$, using different models.
We note that the energy ratio $E_4^{\text{gr}}/E_3^{\text{gr}}$
of $9.74$
(see Table~\ref{table_modelII})
for the $\kappa=1$ system with large repulsive three-body force
is about 20~\% smaller than the energy 
ratio $E_4^{\text{gr}}/E_3^{\text{gr}}$ of $12.21$
obtained in the absence of the three-body force
(see Table~\ref{table_modelI}).
This indicates that the $\kappa=1$ results reported in Table~\ref{table_modelI}
are not universal despite the fact that the ratio $|E_3^{\text{gr}}|/E_{\text{sr}}$
is rather small.

Interestingly, Ref.~\cite{PRXMonash} reported the existence 
of an extremely weakly-bound excited four-body state
for $\kappa=1$ (see Table~I of Ref.~\cite{PRXMonash}).
For our Model~II, we were not able to find such a state.
Looking ahead, we note
that our calculations 
for Model~III with large
$d_{\text{BX}}$ suggest, in agreement with
Table~\ref{table_modelII}, that the $\kappa=1$ and $12$ systems 
do not support an excited four-body state at unitarity.
While we cannot rule out that this
is due to the variational character of
our calculations (i.e., an excited state is supported by Model~II
but we missed it),
we speculate that the disagreement between our results and Ref.~\cite{PRXMonash} 
points toward a sensitive dependence
of the energy ratios on the underlying model interaction.

Table~\ref{table_modelII} also reports five-body 
energies.
These will be discussed in more detail in the next section.

\begin{widetext}

\begin{table}
\caption{Energies of the B$_{N-1}$X system, $N=3-4$, interacting through
a BX Gaussian potential with infinitely large
$s$-wave scattering length $a_{\text{BX}}$, 
repulsive three-body Gaussian potential,
and vanishing BB potential ($d_{\text{BB}}=0$)
for various mass ratios
(Model~II).
The $\kappa=1$, $2$, and $4$ energies are obtained for
$d_{\text{BBX}}=4.8 E_{\text{sr}}$; the reported energies should be close
to the $d_{\text{BBX}} \rightarrow \infty$ limit.
We find $|E_3^{\text{gr}}|/E_{\text{sr}}=2.4 \times 10^{-8}$,
$3.1 \times 10^{-6}$, and $4.6 \times 10^{-5}$ for $\kappa=1$, $2$, and $4$, respectively.
The entry ``$-$'' indicates that a bound state was not found.
The energies for $\kappa=8-50$
are
taken from Ref.~\cite{BlumeYanPRL}.
}
\begin{ruledtabular}
\begin{tabular}{l|llll}
$\kappa$ 
& $E_3^{\text{gr}}/E_3^{\text{exc,1}}$ & 
$E_4^{\text{gr}}/E_3^{\text{gr}}$ &  $E_4^{\text{exc,1}}/E_3^{\text{gr}}$ &  
$E_5^{\text{gr}}/E_3^{\text{gr}}$  \\
\hline
$1$ 
& & $9.51$ & $-$  & $25.1$  \\
$2$ 
 & $$ & $4.85$ & $-$ & $9.74$ \\
$4$ 
 & $$ & $3.36$ & $-$  & $5.71$  \\
$8$ 
 & $(12.510)^2 \approx 156.5$ & $(1.647)^2 \approx2.713$ & $-$ & $(2.06)^2 \approx4.244$  \\
$12$ 
 & $(8.158)^2 \approx66.55$ & $(1.58)^2 \approx2.496$ & $-$ & $(1.94)^2 \approx3.764$  \\
$16$ 
 & $(6.313)^2 \approx39.85$ & $(1.544)^2 \approx2.384$ & $(1.002)^2 \approx1.004$ & $(1.88)^2 \approx3.534$  \\
$133/6$ 
 & $(4.904)^2 \approx24.05$ & $(1.510)^2 \approx2.280$ & $(1.010)^2 \approx1.020$ & $(1.82)^2 \approx3.312$  \\
$30$ 
 & $(3.998)^2 \approx15.98$ & $(1.488)^2 \approx2.214$ & $(1.026)^2 \approx1.053$ & $(1.78)^2 \approx3.168$  \\
$40$ 
 & $(3.372)^2 \approx11.37$ & $(1.471)^2 \approx2.164$ & $(1.046)^2 \approx1.094$ & $(1.75)^2 \approx3.063$  \\
$50$ 
& $(2.996)^2 \approx8.714$ & $(1.461)^2 \approx2.135$ & $(1.067)^2 \approx1.138$ & $(1.73)^2 \approx2.993$  
\end{tabular}
\end{ruledtabular}
\label{table_modelII}
\end{table}

\end{widetext}

\subsection{Model~III}

Reference~\cite{GiorginiPRA1} investigated the equal-mass polaron 
problem by modeling the boson-boson interaction by a
repulsive two-body step potential.
It was later argued~\cite{PRXMonash} that the results for the interaction
model used in Ref.~\cite{GiorginiPRA1}
(basically,
our Model~III with repulsive and attractive two-body step potentials
instead of repulsive and attractive two-body
Gaussian potentials)
should be universal, provided the energies are scaled by
the trimer ground state energy.
Interestingly, Ref.~\cite{GiorginiPRA1} found four- and five-body
bound states but no six-body bound state.
It was suggested~\cite{PRXMonash} that this may be due to
the fact that a single impurity can bind one $s$-
and three $p$-wave bosons and that shell
closure prevents the binding of additional bosons.
In the following, the question of universality and
the existence of six-body bound states 
is investigated using Model~III.

\begin{figure}
\vspace*{+.9cm}
\includegraphics[angle=0,width=70mm]{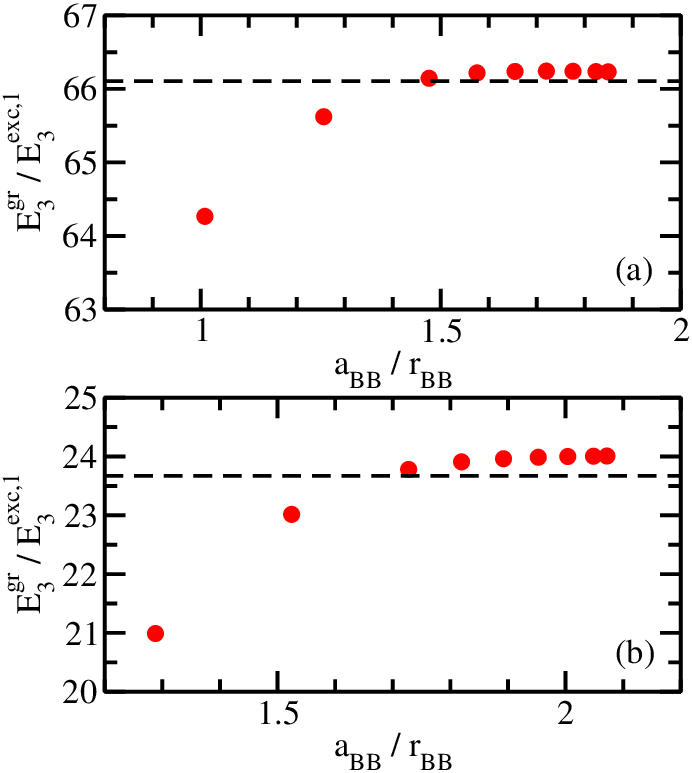}
\vspace*{0.2cm}
\caption{(color online)
Characteristics of Model~III at unitarity.
Circles show the energy
ratio $E_3^{\text{gr}}/E_3^{\text{exc},1}$ 
as a function of $a_{\text{BB}}/r_{\text{BB}}$ for 
(a) $\kappa=12$ and (b) $\kappa=133/6$.
For comparison, dashed horizontal lines show the zero-range scaling 
parameter
$\lambda^2$ from Table~\ref{table_zr}.
}
\label{fig_energybb1}
\end{figure}

Circles in Figs.~\ref{fig_energybb1}(a)  and
\ref{fig_energybb1}(b) 
show the 
energy ratio $E_3^{\text{gr}}/E_3^{\text{exc},1}$ as a function
of $a_{\text{BB}}/r_{\text{BB}}$ for $\kappa=12$ and $\kappa=133/6$,
respectively.
For comparison, the horizontal dashed lines show the
scaling parameter $\lambda^2$ from the zero-range theory,
which assumes vanishing BB interactions.
The energy ratios for Model~III plateau 
as $a_{\text{BB}}/r_{\text{BB}}$
increases at a value somewhat larger 
than that predicted by the 
zero-range theory.
The deviation between
$E_3^{\text{gr}}/E_3^{\text{exc},1}$ 
for the largest $a_{\text{BB}}$ considered
and the zero-range scaling parameter is
0.19~\% and 1.4~\% 
for $\kappa=12$ and $\kappa=133/6$, 
respectively.
This 
shows that the actual
value of the $a_{\text{BB}}$ scattering length
plays a secondary role, provided $a_{\text{BB}}$ is much smaller than the
size of the ground state trimer.
A sufficiently large positive value of $a_{\text{BB}}$
leads to the exclusion
of a portion of the configuration
space,
thereby bringing the results closer to the universal regime.
We were unfortunately not able
to reliably determine $E_3^{\text{exc},1}$ for $\kappa=1$
due to the extremely large scaling parameter. 
We expect 
that $E_3^{\text{gr}}/E_3^{\text{exc},1}$ 
would reach a plateau for smaller $a_{\text{BB}}/r_{\text{BB}}$ 
than for $\kappa=12$ and that the percentage
deviation between the plateau value
and the zero-range scaling parameter would be smaller than
the percentage deviation for $\kappa=12$.

\begin{figure}
\vspace*{+.9cm}
\includegraphics[angle=0,width=70mm]{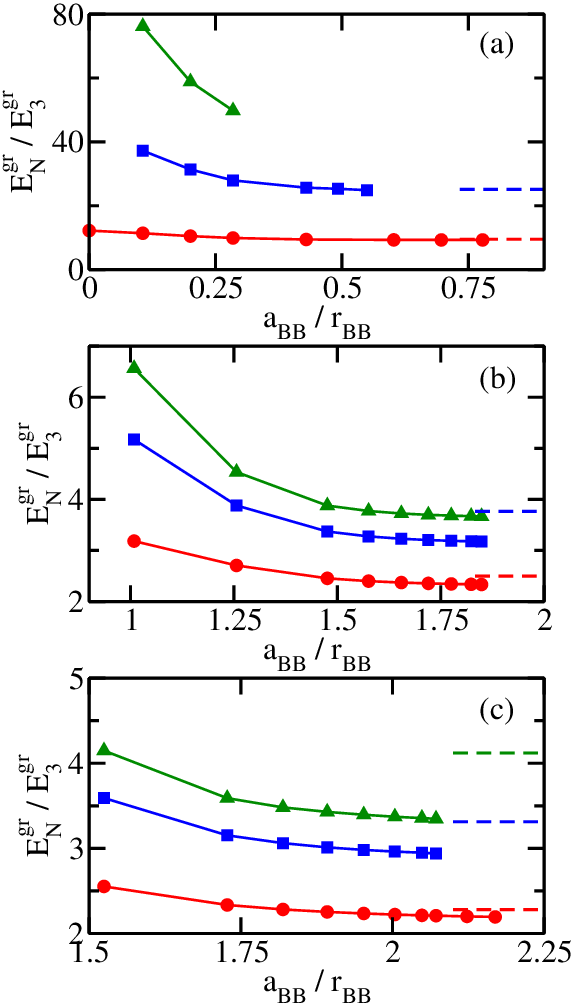}
\vspace*{0.2cm}
\caption{(color online)
Comparison of ground state energies
for Models~II and III at unitarity.
Circles, squares, and triangles show the energy
ratios $E_4^{\text{gr}}/E_3^{\text{gr}}$,
$E_5^{\text{gr}}/E_3^{\text{gr}}$, and
$E_6^{\text{gr}}/E_3^{\text{gr}}$, respectively,
as a function of $a_{\text{BB}}/r_{\text{BB}}$ for (a) $\kappa=1$,
(b) $\kappa=12$,
and
(c) $\kappa=133/6$ for Model~III;
the lines connecting the symbols
serve as a guide to the eye. 
The dashed horizontal lines show the 
energy ratios
for Model~II (see Table~\ref{table_modelII}).
The $N=6$ energy ratio for Model~II is only shown for $\kappa=133/6$.
}
\label{fig_energybb2}
\end{figure}

Figures~\ref{fig_energybb2}(a)-\ref{fig_energybb2}(c)
summarize our $N=4-6$ results for $\kappa=1$, $\kappa=12$, and $\kappa=133/6$,
respectively.
For all three mass ratios, the change of
$E_N^{\text{gr}}/E_3^{\text{gr}}$ decreases with increasing 
$a_{\text{BB}}/r_{\text{BB}}$.
For fixed mass ratio $\kappa$, the energy ratio 
$E_4^{\text{gr}}/E_3^{\text{gr}}$ (circles) reaches a plateau quicker than the 
energy ratios $E_5^{\text{gr}}/E_3^{\text{gr}}$ (squares)
and $E_6^{\text{gr}}/E_3^{\text{gr}}$ (triangles).
Also, the ``flattening'' with increasing $a_{\text{BB}}/r_{\text{BB}}$
is
faster for $\kappa=1$ than for $\kappa=12$ and $133/6$.
For comparison, the dashed horizontal lines
on the right edge of Figs.~\ref{fig_energybb2}(a)-\ref{fig_energybb2}(c)
show the 
energy ratios
for Model~II (see Table~\ref{table_modelII} and Ref.~\cite{BlumeYanPRL}).
For $N=4$, 
the energy ratios for Model~III (circles)
for the largest $a_{\text{BB}}/r_{\text{BB}}$ considered
and Model~II (lowest dashed line)
differ by about 2.4~\%, 7.1~\%, and 3.9~\%
for $\kappa=1$, $12$, and $133/6$, respectively.
The calculations underline that it is challenging
to reach the fully universal regime for large mass ratios by adding purely
repulsive two- or three-body potentials.
The deviations for $N=5$ are 
$0$~\%, 19~\%, and 13~\%
for $\kappa=1$, $12$, and $133/6$, respectively.
Generally speaking, we 
expect that the discrepancy between the two sets of results
would increase with increasing $N$.
For $\kappa=133/6$, this is indeed the case [see Fig.~\ref{fig_energybb2}(c)].
For
$N=6$ and $\kappa=1$, Fig.~\ref{fig_energybb2}(a) shows converged energy ratios up to
$a_{\text{BB}}/r_{\text{BB}} \approx 0.285$;
for larger $a_{\text{BB}}/r_{\text{BB}}$, our energy ratios (not shown) are not fully converged.
For $a_{\text{BB}}/r_{\text{BB}} \approx 0.6$, e.g., we find
$E_6^{\text{gr}}/E_3^{\text{gr}}=35.2$, which should be considered as a lower
bound since our calculations are variational and $E_3^{\text{gr}}$ is highly
accurate.
We conclude that Model~III suports a six-body bound state
in the large $d_{\text{BBX}}$ limit.
Such a bound state was not found in Ref.~\cite{GiorginiPRA1}  
for the square-well model.
This suggests that the shell-closure argument put forward in Ref.~\cite{PRXMonash}
does not hold, in general, for bosonic systems with an impurity.
The discussion surrounding Fig.~\ref{fig_energybb2} can be
summarized as follows:
While Models~II and III predict somewhat different energy ratios
for $N \ge 5$, we believe that these models provide a realistic description
of the hierarchy of few-body states
of small bosonic systems with a single impurity.

\begin{figure}
\vspace*{+.9cm}
\includegraphics[angle=0,width=70mm]{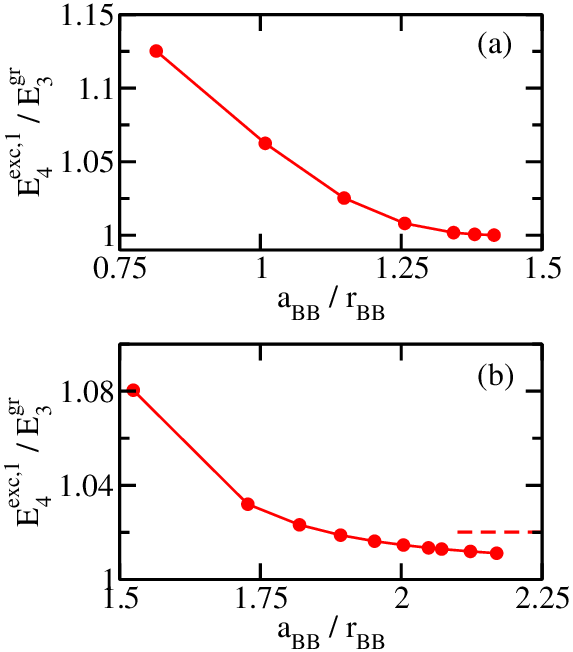}
\vspace*{0.2cm}
\caption{(color online)
Energies of the excited four-body states
at unitarity.
Circles show the energy
ratio $E_4^{\text{exc},1}/E_3^{\text{gr}}$
for Model~III 
as a function of $a_{\text{BB}}/r_{\text{BB}}$ for 
(a) $\kappa=12$
and
(b) $\kappa=133/6$; the lines connecting the symbols
serve as a guide to the eye. 
The horizontal dashed line in panel~(b) shows the 
corresponding energy ratio
for Model~II.
}
\label{fig_energybb3}
\end{figure}

Reference~\cite{BlumeYanPRL} (see also Table~\ref{table_modelII})
found 
that Model~II  at unitarity supports 
an excited four-body state for $\kappa=16-50$.
For $\kappa=12$, in contrast, no such
state was found.
The corresponding results for Model~III are summarized
in Fig.~\ref{fig_energybb3}.
The change of the energy ratio
$E_4^{\text{exc},1}/E_3^{\text{gr}}$ 
decreases as $a_{\text{BB}}/r_{\text{BB}}$ increases.
For the largest $a_{\text{BB}}/r_{\text{BB}}$ considered,
the energy ratio
$E_4^{\text{exc},1}/E_3^{\text{gr}}$ for $\kappa=133/6$ takes 
a value of around $1.011$,
which is somewhat smaller, accounting for error bars, than 
the corresponding value of $1.020$ 
for
Model~2 (according to Ref.~\cite{BlumeYanPRL}, the errorbar is
$0.005$ 
on the square root of the energy ratio).

In agreement with the Model~II results, we find that the
excited four-body state for $\kappa=12$ disappears as 
$a_{\text{BB}}/r_{\text{BB}}$ goes beyond a critical value
[$a_{\text{BB}}/r_{\text{BB}} \gtrsim 1.42$; see Fig.~\ref{fig_energybb3}(a)],
which is smaller than the value for which we would
expect, based on the ground state results shown in Fig.~\ref{fig_energybb2}, 
the energy ratio to be independent of $a_{\text{BB}}/r_{\text{BB}}$.
Figure~\ref{fig_energybb3} indicates that the 
predictions of Models~II and III for the energy ratio
$E_4^{\text{exc},1}/E_3^{\text{gr}}$ are reasonably consistent.

For $\kappa=1$, we find that the excited four-body state
supported by Model~I disappears when the boson-boson
scattering length is sufficiently repulsive (Model~III).
The absense of an excited four-body state at unitarity
for $\kappa=1$, as predicted by Models~II and III, is in disagreement
with the predictions of the ``$r_0$ and $\Lambda$ models'' of Ref.~\cite{PRXMonash}.
In those models, 
an energy for the excited
four-body state, expressed in terms of the 
three-body ground state energy, of $1.0030(3)$
and $1.0036(1)$ was reported for the $r_0$ and $\Lambda$ models, respectively.
Since the binding energy is extremely small,
it might be that a small change in the interaction
model  moves the critical scattering length of the excited
four-body state from the positive to the negative scattering length side, thereby
explaining the discrepancy. Alternatively, it could be that our model
supports an excited four-body bound state at unitarity but
that our numerical approach missed the state.

\section{Results: Negative scattering length side}
\label{sec_negativeas}

This section discusses the behavior of the BBBX system 
with $\kappa=133/6$ as a function of 
the interspecies $s$-wave scattering length $a_{\text{BX}}$
($a_{\text{BX}}<0$).
For Model~II,
the critical BX scattering lengths 
$a_{4,-}^{\text{gr}}$
and 
$a_{4,-}^{\text{exc,1}}$,
at which the ground and first excited tetramer energies are
resonant with the four-atom threshold,
were predicted to be
$a_{4,-}^{\text{gr}} \approx 0.55 a_{3,-}$ and
$a_{4,-}^{\text{exc}} \approx 0.91 a_{3,-}$, respectively~\cite{BlumeYanPRL}.
Here, $a_{3,-}$ is the critical BX scattering length at which
the trimer that the four-body states are tied to becomes unbound
on the negative scattering length side.

\begin{figure}
\vspace*{+.9cm}
\includegraphics[angle=0,width=70mm]{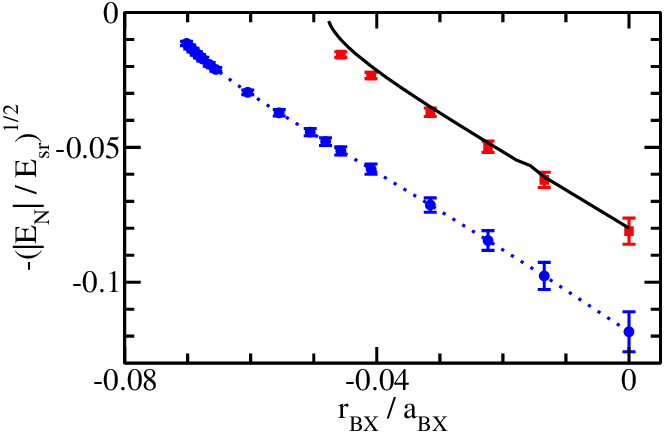}
\vspace*{0.4cm}
\caption{(color online)
Generalized Efimov spectrum, obtained using Model~I, 
for $\kappa=133/6$ as a function
of the inverse of the
BX $s$-wave scattering length $a_{\text{BX}}$ 
(only the negative $a_{\text{BX}}$ regime is shown).
The energies and scattering lengths are scaled by the short-range
quantities $E_{\text{sr}}$ and $r_{\text{BX}}$, respectively
(the energy is shown on a square-root scale).
The solid line shows the energy $E_{3}^{\text{exc},1}$ 
of the BBX system.
The circles and squares show the real part $E_r$ of the
resonance energies of the BBBX system 
(to guide the eye, the dotted line connects consecutive points for
the energetically lower-lying resonance).
The error bars indicate the absolute value of the
imaginary part $E_i$ of the resonance energy
(again, using the square-root scaling);
a smaller $|E_i|/E_{\text{sr}}$ corresponds to a larger scaled lifetime.
}
\label{fig_negativeas}
\end{figure}

To obtain a sense for the dependence of these results
on the underlying model, we additionally performed calculations for the negative
$a_{\text{BX}}$ regime using Model~I.
Specifically, we determine the energy of the four-body
resonances tied to the first excited BBX trimer.
The results are summarized in Fig.~\ref{fig_negativeas}.
The solid line shows the energy $E_3^{\text{exc},1}$
of the first excited trimer while
the circles and squares show the real part $E_r$
of the energetically lower- and higher-lying four-body resonances.
The absolute value $|E_i|$ of the imaginary part is shown by errorbars.
For example, the magnitude of the imaginary
part $E_i$ of the resonance energy of
the energetically lower-lying four-body state
changes from $9 \times 10^{-3} E_{3}^{\text{exc},1}$
at unitarity to around $10^{-4} E_{3}^{\text{exc},1}$
for the point closest to threshold in Fig.~\ref{fig_negativeas}.
The magnitude of the imaginary
part $E_i$ of the resonance energy of
the energetically higher-lying four-body state
changes from $4 \times 10^{-3} E_{3}^{\text{exc},1}$
at unitarity to around $2 \times 10^{-4} E_{3}^{\text{exc},1}$
for the point closest to threshold in Fig.~\ref{fig_negativeas}.
Extrapolating the four-body resonance energies $E_r$ to zero, 
we estimate the
following critical scattering lengths for the two four-body states:
 $a_{4,-}^{\text{gr}} \approx 0.66 a_{3,-}$
and
$a_{4,-}^{\text{exc,1}}  \approx 0.94 a_{3,-}$.
The agreement with the results for Model~II is quite good,
especially considering that the critical scattering lengths have a few percent 
uncertainty due to numerical inaccuracies 
and that
the determination of the critical scattering lengths requires an
extrapolation to the threshold.

\section{Summarizing remarks}
\label{sec_conclusion}

This work determined the bound state energies
of an impurity that interacts with $N-1$ bosonic atoms
through short-range interactions
that are characterized by the $s$-wave scattering length $a_{\text{BX}}$.
For the cases considered, the impurity mass was the same as
or smaller than that of the bosons.
Impurity problems are ubiquitous in
physics, ranging from impurities
in condensed matter systems to impurities in quantum liquid droplets, such
as helium and molecular hydrogen clusters, to
impurities in cold fermionic and bosonic atomic gases.
A key objective of the present
work was to investigate, using two-body interactions that mimick
zero-range interactions in the limit that the trimer size is large compared to
the range of the two-body potential, universal four- and
higher-body states 
that are linked
to three-body Efimov trimers consisting of two bosonic atoms and the impurity.
To address this objective, the results for different interaction models
were compared. While the present work considered two-body
single-channel models, Ref.~\cite{PRXMonash} treated the $N=4$ system
with $\kappa=1$ using a two-body coupled-channel model.

The impurity problem studied in this work is unique due to
its close connection to three-body Efimov states.
In the large $|a_{\text{BX}}|$ limit, the
weakly-bound BBX states follow Efimov's radial scaling law,
which implies that the three-body states are governed
by the $s$-wave scattering length and
a three-body parameter.
If the four-body states are fully governed by these parameters,
then different interaction models should, in the limit
that the effective range corrections can be neglected,
yield the same value for the four-body energies, provided
the four-body energies are expressed in terms of the three-body energy
and provided the $s$-wave scattering lengths are the same.
This work shows that this is the case
for $N=4$ and $\kappa=1$. As the mass ratio $\kappa$ increases,
model dependencies at the few percent level develop.
For the five-body system, the energy ratio $E_{5}^{\text{gr}}/E_3^{\text{gr}}$ 
displays, for $\kappa=12$ and $133/6$, a stronger model dependence than the 
energy ratio $E_{4}^{\text{gr}}/E_3^{\text{gr}}$.
In general, the ``universality window'' decreases with increasing
number of particles since the binding
energy increases (i.e., the system size shrinks with
increasing $N$). This is particularly prominent when
$\kappa$ is notably larger than $1$.
It would be interesting to extend the very recent
effective field theory study for identical bosons~\cite{EFT} 
to the bosonic system with impurity considered in this work.
Specifically, it would be interesting to explore at which order the
four-body parameter enters.
It should be kept in mind
that the numerical calculations become more challenging
as $N$ increases, implying that it is harder
to exhaustively explore the parameter space
of the model interactions with high accuracy
for $N=5$ and $N=6$.
Model~III suggests, in contrast to what was found in
Ref.~\cite{GiorginiPRA1} for a slightly different model, 
that the system for $\kappa=1$
supports a six-body bound state.
It will be interesting to explore the implications of this
bound state on the physics of the Bose polaron.


\section{Acknowledgement}
Support by the National Science Foundation through
grant number
PHY-1806259
is gratefully acknowledged.

\appendix

\section{Illustration of complex scaling approach}
\label{appendix_numerics}

This appendix 
illustrates the complex scaling approach,
using a basis set constructed from explicitly correlated Gaussian basis functions,
for the BBBX system
with mass ratio $\kappa=8$
for infinitely large BX scattering length,
$d_{\text{BB}}=d_{\text{BBX}}=0$
(Model~I), and $E_{\text{target}}=-2.52 \times 10^{-4}E_{\text{sr}}$.
This target energy is about three times
less negative than the resonance
energy
of $E_{r}=-7.69 \times 10^{-4} E_{\text{sr}}$
reported
in Table~\ref{table_modelIres}.

\begin{figure}
\vspace*{+1.2cm}
\includegraphics[angle=0,width=70mm]{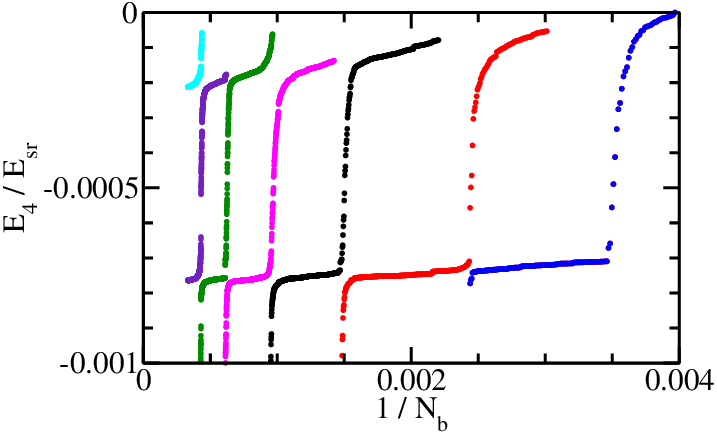}
\vspace*{0.4cm}
\caption{(color online)
Basis set generation for BBBX system
with infinitely large interspecies $s$-wave scattering length
and $\kappa=8$
(Model~I).
The target energy $E_{\text{target}}$ is set to $-2.52 \times 10^{-4} E_{\text{sr}}$.
The symbols show the four-body energies $E_4$ 
as a function of the inverse of the number $N_b$ of basis functions.
The displayed energies correspond to the 8-th through 14-th
eigen value of the generalized eigen value problem.
The ``plateau'' at $E_4 \approx -7.8 \times 10^{-4} E_{\text{sr}}$
is indicative of a four-body resonance
(see also Fig.~\protect\ref{fig_complexscaling}).
The data suggest that there may exist another resonance
at $E_r \approx -2 \times 10^{-4} E_{\text{sr}}$.
However, the complex scaling did not reveal such a resonance.
This is likely due to the fact that the 
energetically higher-lying plateau is not
a signature of a resonance 
but related to a BBX bound state.
}
\label{fig_basisset}
\end{figure}

\begin{figure}
\vspace*{+.5cm}
\includegraphics[angle=0,width=80mm]{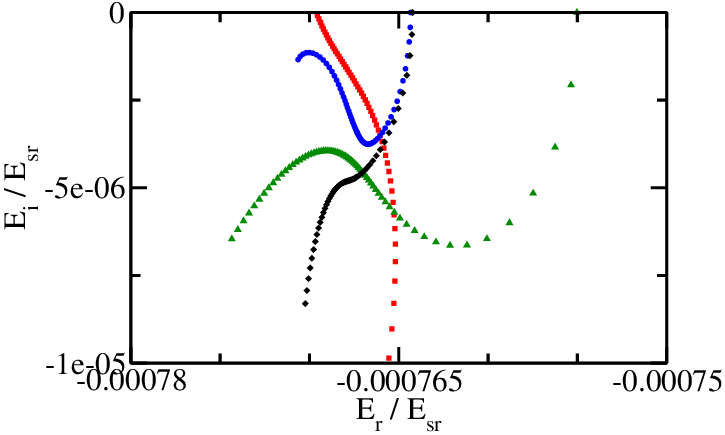}
\vspace*{0.4cm}
\caption{(color online)
Complex scaling
results for the BBBX system
with infinitely large interspecies $s$-wave scattering length
and $\kappa=8$
(Model~I).
Red squares, blue circles, green triangles, and 
black diamonds
show trajectories, generated by scanning the rotation angle $\theta$,
in the 
complex energy plane for $N_b=1600$, $2000$, $2500$,
and $3000$, respectively
($\theta$ is increased linearly in steps
of approximately
$6.957 \times 10^{-3}$radians for $N_b=1600$,
$8.276 \times 10^{-3}$radians for $N_b=2000$,
$6.667 \times 10^{-3}$radians for $N_b=2500$, and
$8.421 \times 10^{-3}$radians for $N_b=3000$).
For vanishing rotation angle, the imaginary 
part of the energy is zero.
This figure and Fig.~\ref{fig_basisset} are obtained
using the same
basis set.
}
\label{fig_complexscaling}
\end{figure}

To illustrate the construction of the basis set,
Fig.~\ref{fig_basisset} shows 
the eigen values as
a function of the inverse of the number $N_b$ of basis functions.
For the example at hand, the first 100 basis functions were chosen
such that the two bound four-body states 
(see Table~\ref{table_modelI}) are reasonably well described.
For
$N_b=250$ (right edge of the figure), the state
with energy larger than and closest to the target energy
corresponds to the 8-th eigen value.
As more basis functions are added, the energy of the 8-th
state drops below $E_{\text{target}}$ (this occurs
around $1/N_b=0.0035$ in Fig.~\ref{fig_basisset})
and the next higher-lying state
is being optimized. This ``dropping down'' is repeated several times
during the optimization procedure.
The reason that the energy ``drops'' during the optimization is
that there exists a continuum of ``trimer-plus-atom states''
above the  three-body ground state. Since the basis functions
have a finite as opposed to an infinite spatial extend
and since the basis set is finite, the
continuum is discretized. The roughly flat portion (plateau
at $E_4 \approx -7.7 \times 10^{-4} E_{\text{sr}}$) of the 
eigen values is, as confirmed by the results presented
in Fig.~\ref{fig_complexscaling}, associated with a resonance state.

To extract quantitative information, 
we solve the eigen value problem spanned by $\tilde{\underline{H}}_{\text{rel}}$
and $\underline{O}$ for various $\theta$.
The resulting eigen values are categorized as 
corresponding to bound states, scattering states,
and resonance states according to the
behavior of the  trajectories in the 
complex plane. To locate the resonance states,
we plot the trajectories, which span several orders of magnitude
in $E_r$ and $E_i$, in different energy windows.
Squares, circles, triangles, and diamonds
in Fig.~\ref{fig_complexscaling} show trajectories
corresponding to a resonance
state using basis sets
with $N_b=1600$, $2000$, $2500$, and $3000$ 
(the same basis functions as used in Fig.~\ref{fig_basisset}).
The ``beginning point'' ($\theta=0$) of the trajectories 
can be identified by the condition
that  the imaginary part of the energy is zero for $\theta=0$.
For each trajectory, the symbols are obtained for equally spaced
$\theta$.
It can be seen that the trajectories for the
different basis set sizes all go, roughly, through the
point
$(E_r,E_i)\approx
(-7.69 \times 10^{-4} E_{\text{sr}},-5.3 \times 10^{-6} E_{\text{sr}})$.
Moreover, at or near this point in the complex energy plane, the 
trajectories
for $N_b=2000$, $2500$, and $3000$
slow down; this can be seen from the decreased spacing
of the symbols.
For the example shown,
the calculations for $N_b=1600$ do not allow us to extract the
resonance energy and lifetime.
It is ensuring, though, that the results for $N_b=2000-3000$
agree with each other. 
Repeating the calculations for different $E_{\text{target}}$
to ensure independence of $E_{\text{target}}$,
we extract the resonance position and its lifetime.
The resonance energy moves somewhat for different $E_{\text{target}}$
and different basis sets.
The results reported in Table~\ref{table_modelIres} are,
in most cases, averages from multiple runs.

In general, we find that the resonance position 
(i.e., the real part $E_r$) is numerically more stable 
than the lifetime $\tau$
[which is proportional to the inverse of the imaginary part, $\tau=h/(2|E_i|)$].
Also, as a rule of thumb, the closer the real part
$E_r$ is to zero, the harder
it is to reliably extract the lifetime from our calculations.
Because of this, our complex scaling calculations 
(see Table~\ref{table_modelIres}) are restricted to mass ratios
$\kappa \ge 4$).

\end{document}